\documentstyle[11pt,binaries,twoside,epsf]{article}
\reversemarginpar

\begin{document}
\title{Multiplicity of massive stars}
 \author{Thomas Preibisch, Gerd Weigelt}
\affil{Max-Planck-Institut f\"ur Radioastronomie, Auf dem H\"ugel 69, 
D--53121 Bonn, Germany
}
\author{Hans Zinnecker}
\affil{Astrophysikalisches Institut Potsdam, An der Sternwarte 16,
D--14482 Potsdam, Germany}

\begin{abstract}
We discuss the observed multiplicity of massive stars and
implications on theories of massive star formation.
After a short summary of the literature on massive star multiplicity,
we focus on the O- and B-type stars in the Orion Nebula Cluster,
which constitute a homogenous sample of very young massive stars.
13 of these stars have recently been the targets of a bispectrum
speckle interferometry survey for companions.
Considering the visual and also the known
spectroscopic companions of these stars, 
the total number of companions is at least 14.
Extrapolation with correction for the unresolved systems
suggests that there are {\em at least 1.5 and perhaps as much as 4
 companions per primary star on
average}. This number  is clearly higher than the mean number of $\sim0.5$
companions per primary star
found for the low-mass stars in the general field population and also
in the Orion Nebula cluster.
This suggests that a different mechanism is at 
work in the formation of high-mass multiple systems in the dense Orion Nebula 
cluster than for low-mass stars.
\end{abstract}


\section{Introduction}

It is a well known fact that many, perhaps most
stars in our galaxy are members of multiple systems.
Although this paper is supposed to deal with massive stars
$(M_\ast \ge 8 - 10\,M_\odot$), we will
start with a brief look at the multiplicity of solar type stars,
first because this is very well known,
and second because it is important  to compare
the multiplicity of the massive stars to that of the solar type stars.
The main characteristics of the multiplicity of solar type field stars
can be summarized as follows (c.f.~Duquennoy \& Mayor 1991):
The binary frequency (i.e.~the probability that a given object is multiple;
cf.~Reipurth \& Zinnecker 1993) is about 45\%,  and
any primary star has about 0.5 companions on average.
The ratio of
single\,:\,double\,:\,triple\,:\,quadruple  systems is
 57\,:\,38\,:\,4\,:\,1. The 
distribution of orbital periods follows a log--normal
relationship, and the
mass ratio distribution decreases for
q := M$_2$/M$_1$ $\rightarrow$ 1 and seems to be
consistent with random pairing from the field star initial mass function.

There are several reasons why the multiplicity of massive stars is less 
well known than for the low-mass stars. One of them is that
massive stars are far less numerous than low-mass stars, and thus
they are typically located at larger distances. Another problem is 
caused by the very high luminosities of massive stars, causing
extreme brightness contrasts between the massive primary star and 
any low-mass companion, strongly hampering the detection of visual companions.
However, since a good knowledge of the multiplicity of massive stars can
yield very important constraints on high-mass stellar formation theories
(see below), a study of the multiplicity of massive stars is highly desirable.

\section{Formation of massive stars}

The formation mechanism of massive stars
is still not well understood (cf.~Stahler et al.~1999).
In contrast to low- and intermediate-mass stars, high-mass stars cannot form
through gravitational collapse in molecular cloud cores and subsequent
accretion, because, as soon as
the stellar core reaches a mass of $\sim 10\,M_\odot$, the radiation pressure
on the infalling dust halts the accretion and thus limits the mass
(Yorke \& Kr\"ugel 1977; Yorke 1993).

An alternative model for the formation of massive stars is based on
collisions and subsequent mergers of low- or intermediate-mass protostars
in the dense centers of young clusters.
At first sight, the typical conditions in star forming regions
(stellar number densities of $n_\ast \la 10^{5}$ stars pc$^ {-3}$
and velocity dispersions of $\sigma_v  \approx$ 2 km/sec)
suggest extremely long collision time scales of 
$t_{\rm coll} \ga 10^{10}$ years. This seems to indicate that such
models are not very reasonable.
However, in a very young cluster 
there are important peculiarities that can dramatically change the picture.
One aspect is that very young stars are typically
surrounded by  circumstellar
disks or envelopes, making their geometrical cross section much larger
than the stellar cross section. Another aspect is that the 
cross section can be strongly enhanced in moderately slow encounters
by the gravitational focusing effect.
Furthermore, it is important to take into account that during gravitational
interactions stars do not behave like solid bodies:
close stellar encounters (pericenter $\la 4$ stellar radii)
 can induce huge tidal waves on the stars, 
dissipating kinetic energy into tidal heat; this eventually can lead to the
formation of bound systems or even stellar mergers (cf.~Mardling 1995).

Another crucial factor is that the forming star cluster not only contains 
young stars, but also
an appreciable amount of dense gas, from which the forming stars
accrete mass. 
Bonnell et al.~(1998) have performed theoretical investigations 
and showed that the transfer of mass from diffuse gas to stars
decreases the total energy of the cluster and also the
stellar velocity dispersion. This leads to a strong shrinking of the
cluster, which strongly boosts the stellar number density.
Altogether these
effects can cause the collision time scale to greatly decrease 
to values as low as $t_{\rm coll} \approx 10^{5}$ years, making
the collision and subsequent merger of low- or intermediate-mass protostars
an apparently viable way to form  massive stars, especially in the
center of dense clusters.

If this mechanism is indeed the way how the majority of massive stars form, 
one would expect multiple systems to be 
very common amongst the massive stars, because many of the collisions
will not lead to stellar mergers but to the formation of multiple systems.
Observational multiplicity studies in very young clusters thus can 
help to test this theory.

\section{Observational results on OB star multiplicity}

\subsection{Summary of literature results}

Many previous studies have investigated the
multiplicity of massive stars. In the following short compilation,
which is by no means meant to be complete, we summarize a few of the
most important results.

Abt (1979)  
studied the multiplicity of early B-type (B2-B5) stars. 
By applying a correction for observational incompleteness 
he estimated that the
true binary frequency among these stars might be as high as
$\approx$ 65\%.
He also found that the distribution of orbital periods
is similar to 
that of solar-type binaries, and concluded that the
binary characteristics of normal stars do not depend sensitively
upon primary mass.

In a review paper, Abt (1983) concluded that the spectroscopic
binary frequency among
B-type stars seems to be higher than among F- and G-type stars.
In a further study,
Abt et al.~(1990) 
searched for spectroscopic binaries among 116 B2-B5 stars.
This study indicated that the number of companions is rather high:
assuming that the companion mass function is equal to the Salpeter IMF,
the correction for 
observational incompleteness suggested 
that there are at least 0.8 (1.9) companions with $M_2 \ge 2\,M_\odot$
($M_2 \ge 1\,M_\odot$) per primary star on average. This would be much 
higher than the mean number of 0.5 companions per solar type field star
primary.

Morrell \& Levato (1991) 
spectroscopically studied 96 OB stars in the Orion OB1 association
and found that the 
 percentage of spectroscopic binaries with periods of  $P <100$ days is 32\%,
significantly more than among solar type field stars.
Garmany et al.~(1980) 
derived a spectroscopic binary frequency of 36\% among a sample
of 67 O stars. They found that
 85\% of the systems have mass ratios of $q > 0.4$, i.e.~most of the
companions to these high-mass stars are again high-mass stars.

McAlister et al.~(1993) 
performed speckle interferometric observations of a huge
sample of 2088 OB stars, representing 
  23\% of the members of the {\em Bright Star Catalogue}.
Their observed binary frequency of B stars was 14\%.

Mason et al.~(1998) 
performed a
speckle survey among a magnitude limited (V $<$ 8) sample 
of 227 O-type stars. They detected
15 binaries in the range of separations 0.035$''$ $<$ $\rho$ $<$ 1.5$''$.
Taking into account the
 previously known visual and spectroscopic companions, their results
demonstrated that at least
$60\%$ of the O-type stars have companions.
The distribution of mass ratios they derived is flat or decreasing  
for q $\rightarrow$ 1, in a notable contrast to the result of Garmany 
et al.~(1980).

\subsection{General conclusions on OB star multiplicity}

The above mentioned studies agree on the result that the
multiplicity of OB stars is rather high, probably at least as
high as for the solar-type field stars. 
The derived mass ratio distributions, however, differ strongly from
study to study.

Many of the studies mentioned above share a serious disadvantage by 
being based on rather inhomogeneous samples of stars.
Despite the advantage of allowing to investigate a large number
of stars, any magnitude limited sample is a mixture of stars of
different distances, different ages, and different origin.
This will inevitably cause very serious and complicated selection effects.

Therefore, it was our goal to study a more homogeneous sample of 
massive stars. A very good way to do this is to 
study the population of
massive members of a nearby star forming region. In such a sample,
all the stars have a common distance, age, and origin.
Another important factor is that a very young cluster is especially well
suited for the detection of binary companions since any low- or
intermediate mass companion will still be  in its pre-main sequence phase
and thus typically a factor of $\sim 2 - 10$ brighter than on
the main sequence. This significantly decreases
the enormous difference in brightness between the luminous primary star and its
low-mass companion, which usually makes the observational 
detection of the companion very difficult or even impossible.

\section{Multiplicity of the massive stars in the Orion Nebula Cluster}

The Orion Nebula Cluster (ONC) is a very promising target for a study of
the multiplicity of massive stars.
It is nearby (450 pc), very young ($<$ 1 Myr), and its 
high central stellar density 
($n_\ast \sim 5 \times 10^4\,{\rm pc}^{-3}$) suggests that
stellar collisions might actually be important.
 Also, it is very well investigated 
(e.g.~Genzel \& Stutzki 1989; Herbig \& Terndrup 1986;
McCaughrean \& Stauffer 1994; Hillenbrand \& Hartmann 1998)
and the stellar population is well known (cf.~Brown et al.~1994; 
Hillenbrand 1997).
The compilation of Hillenbrand (1997) lists 27 O- and B-type stars 
as ONC members; these stars constitute a homogenous sample of 
very young massive stars.

Several multiplicity studies have been performed for the
{\em low-mass stars} in the ONC (cf.~Padgett et al.~1997;
Petr et al.~1998; Simon et al.~1999). These studies seem to agree
that the multiplicity of the low-mass stars in the ONC is 
consistent with that in the general field, i.e.~there is {\bf no} evidence
for an overabundance of multiple systems as has been found for the
low-mass pre-main sequence stars in the Taurus region
(c.f.~Leinert et al.~1993, 1997;
K\"ohler \& Leinert 1998; Ghez et al.~1997).

As far as the high-mass stars in the ONC are concerned,
a number of searches for spectroscopic binaries have been performed
(e.g.~Abt et al.~1991; Morrell \& Levato 1991). The Trapezium stars
have been observed with near-infrared speckle holographic methods by
Petr et al.~(1998) and with adaptive optics methods by Simon et al.~(1999).
In order to get more complete information about the multiplicity 
of these stars, we have recently performed 
a bispectrum speckle interferometry (cf.~Weigelt 1977) survey 
of 13 of the ONC O- and B-type stars at the SAO 6 m telescope
(Weigelt et al.~1999; Preibisch et al.~1999).
In combination with the information on the spectroscopic companions,
this gives a comprehensive (though, of course, still not 100\% complete)
picture of the multiplicity of these stars.

\subsection{Known companions to the 13 Orion OB stars}

In our speckle images 8 visual companions have been found  in total.
Considering both, the visual and 
the spectroscopic companions of the 13 target stars, 
the total number of companions is at least 14.
The properties of the known companions are summarized in Table 1.

\begin{table}
\caption[]{Summary of all known companions of the observed ONC 
stars. References:
 1: Preibisch et al.~(1999); 2: Weigelt et al.~(1999); 3: Petr et al.~(1998); 4: Bossi et al.~
(1989); 
 5: Simon et al.~(1999); 6: Abt et al.~(1991); 7: Levato \& Abt (1976).}
\vspace{1mm}

\begin{center}
\begin{tabular}{|llr|lrlr|} \hline
\multicolumn{2}{|c}{Primary} & $M_1$ & Companion & $\rho\;\;\;$ & $\;\;\;\;\;\;q$ & Ref \rule[-0mm]{0mm}{4mm} \\ 
Par-& other name& $[M_\odot]$& & [AU] & & \rule[-3mm]{0mm}{7mm} \\ \hline
1605-1&\,\,\,V372 Ori & 3.5 &-2 (spec)&  & $\sim 0.95$& 7\rule[-0mm]{0mm}{4mm}\\ \hline
1744&\,\,\,HD\,36981\,\,\,\,&4.8& $-$&  &  &\rule[-0mm]{0mm}{4mm} \\\hline
1772&\,\,\,LP Ori & 7.2 &  $-$&  &  &\rule[-0mm]{0mm}{4mm}\\\hline
1863-1&\,\,\,$\theta^1{\rm Ori}$\,B\,\,\,\,&  7 &-2 (vis)& 430  & $\sim 0.22\,$ & 1,2\rule[-0mm]{0mm}{4mm}\\
  & &  &-3 (vis)& 460  & $\sim 0.10\,$ & 1,2\\
  & &  &-4 (vis)& 260  & $\sim 0.03\,$ & 1,5\\ 
  & &  &-5 (spec)&0.13 &   & 6\\ \hline
1865-1&\,\,\,$\theta^1{\rm Ori}$\,A & 16 &-2 (vis)& 100  & $\sim 0.25$& 2,3\rule[-0mm]{0mm}{4mm}\\
  & & &-3 (spec)&  1  & $\sim 0.13 $ & 4 \\ \hline
1889&\,\,\,$\theta^1{\rm Ori}$\,D & 17 &  $-$&  &  &\rule[-0mm]{0mm}{4mm}\\\hline
1891-1&\,\,\,$\theta^1{\rm Ori}$\,C & 45 &-2 (vis)&  16  & $\sim 0.12$& 2\rule[-0mm]{0mm}{4mm}\\ \hline
1993-1&\,\,\,$\theta^2{\rm Ori}$\,A & 25 &-2     (vis)&  173  & $\sim 0.28\,$ & 1\rule[-0mm]{0mm}{4mm}\\ 
    &&   &-3     (spec)& 0.47 & $\sim 0.35$ &6\\  \hline
2031&\,\,\,$\theta^2{\rm Ori}$\,B &12 & $-$ &  & &\rule[-0mm]{0mm}{4mm} \\\hline
2074-1&\,\,\,NU Ori & 14 &-2     (vis)& 214  & $\sim 0.07\,$  & 1\rule[-0mm]{0mm}{4mm}\\ 
    &&   &-3     (spec)& 0.35 & $\sim 0.2$ &6\\  \hline

2271-1&\,\,\,HD\,37115 &  5 &-2    (vis)&  400 & $\sim 0.29\,$  & 1\rule[-0mm]{0mm}{4mm}\\  \hline
2366&\,\,\,HD\,37150 & 15 & $-$ &  & &\rule[-0mm]{0mm}{4mm}  \\\hline
2425-1&\,\,\,WH 349 & 4 &-2     (vis)&  388 & $\sim 0.04\,$  & 1\rule[-0mm]{0mm}{4mm}\\  \hline
\end{tabular}  \end{center}
\label{companions}
\end{table}

With 14 known companions to the 13 target stars, the
mean number of observed companions per primary star is 1.1. Although this
clearly is a firm lower limit to the true number of companions,
it is already 2 $\times$ more than the average number of companions
among the low-mass field star primaries.


\subsection{Estimation of the true number of companions}

Obviously, the number of known companions is only a lower limit to
the true number of companions, as any observational search for companions
is subject to limited sensitivity.
Preibisch et al.~(1999) estimated the fraction of companions that might
have been not detected in the speckle observations due to their being
either too close to be resolved ($\rho \la 35$ mas) or too faint to
be detected ($\Delta K \ga 4$ mag).
The correction factor, i.e.~the ratio of all companions to 
detected companions, was estimated to be between 2.5 and 6.7,
depending on the assumptions about the underlying mass ratio distribution
and the distribution of separations.
Since the speckle observations revealed 
8 visual companions to 13 target stars, the extrapolated  
number of companions is probably in the range  $\approx$\,\,[20 \dots\,50].
This suggests that there are 
$\approx$\,\, [1.5 \dots\,4]\,\,  companions per primary star on average,
and this is
at least 3 $\times$ more than for low-mass primaries.

\subsection{Properties of the multiple systems}

\noindent
{\bf \hspace{6mm} $\bullet$\,   Distribution of mass ratios} \smallskip

\noindent
While the mass ratio distribution for low-mass binaries is rather well
known to be consistent with the field IMF,
several very different distributions have been derived for the mass ratios
in different samples of high-mass binary systems. The results range from
distributions favoring low-mass companions (e.g.~Mason et al.~1998),
over flat distributions, to distributions which more or less strongly 
favor relatively massive companions
(e.g.~Abt \& Levy 1978;  Garmany et al.~1980).
In our sample, 
all known companions are low- or intermediate mass stars and the distribution
of mass ratios has a strong peak at moderately low values.

\begin{figure}
\plotone{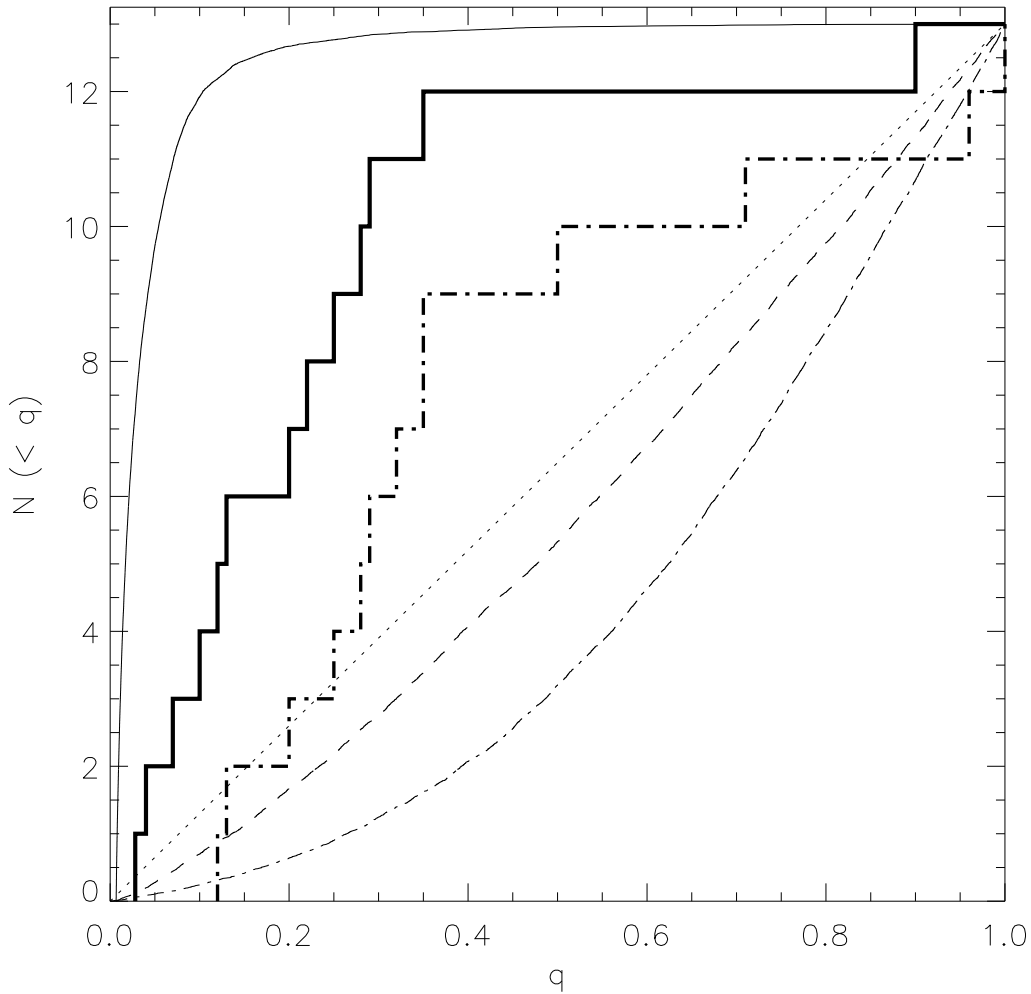}
\caption{The thick solid line shows the cumulative distribution function 
of observed mass ratios of the ONC multiple systems, based on our best
estimates of the companion masses. The thick dashed-dotted line shows the 
same distribution based on very conservative upper mass limits for the
companions (for details see Preibisch et al.~1999).
These empirical distributions are 
compared to the four different model distributions described in the text:
({\bf a)}: random pairing from the Scalo (1998) field star IMF 
 (thin solid line); 
({\bf b)}: flat mass ratio distribution (thin dotted line);
({\bf c)}: a distribution that slightly favors massive companions 
 (dashed line); 
{\bf (d)}: a distribution that strongly favors systems with
(nearly) equal masses (thin dashed-dotted line).
}
\end{figure}

In Fig.~1 we compare our empirical mass ratio distribution with the 
following different models:

\noindent
{\bf (a)} \parbox[t]{12.5cm}{a mass ratio distribution given by random
pairing of stars drawn from the Scalo (1998) field star IMF;}

\noindent
{\bf (b)} \parbox[t]{12.5cm}{a flat distribution of mass ratios, i.e.
$ f(q) = \rm const.$ for $[0 \le q \le 1]$;}

\noindent
{\bf (c)} \parbox[t]{12.5cm}{a distribution that slightly favors massive companions,
$f(q) \; \propto \; q^{0.25}$ for $[0 \le q \le 1]$,  
as derived by Abt \& Levy (1978) for a sample of early B-type stars;}

\noindent
{\bf (d)} \parbox[t]{12.5cm}{a distribution of mass ratios which strongly favors systems with 
(nearly) equal masses. Here we consider the findings of the binary survey among 
O-type stars performed by Garmany et al.~(1980), which can be roughly 
approximated by a half Gaussian distribution with a peak at  
$ q = 1$ and a width of $\sigma \approx 0.45$
for $[0 \le q \le 1]$;}\smallskip

\noindent
It is rather obvious that our empirical distribution is not
consistent with the models (c) and (d).
Thus we conclude that any distribution  favoring high mass ratios
seems to be definitely excluded by our data.
Furthermore, one has to keep in mind that most probably there are more, 
still undetected faint low-$q$ companions, which are missing in our 
empirical distribution. Even if we take into account that our estimates
of the mass ratios are subject to significant uncertainties
(see Preibisch et al.~1999 for a detailed discussion), it appears
very likely that the true mass ratio distribution is significantly peaked 
towards moderately low mass ratios.
\bigskip

\noindent
{\bf \hspace{6mm} $\bullet$\,  Number of components per system}  \medskip

\noindent
Among the solar type field stars only {\bf 5\%} of the systems are
triple or higher-order systems (Duquennoy \& Mayor 1991).
In our sample, however, {\bf 31\%} of the systems have at least 3 components.
The numbers of systems with given order are compared to the corresponding
numbers for solar type field stars in Table 2. 
One can clearly see that
{\bf the fraction of higher-order multiple systems
seems to be enhanced among the OB stars}.
\begin{table}
\caption{Systems with multiplicity
of a given order in our sample and among the solar type field stars.
Data for the solar type field stars are from Duquennoy \& Mayor (1991).}
\begin{center}
\begin{tabular}{|l|cccc|}\hline
& singles & binaries & triples & quadruples\\\hline
number&  5      &   4      &   3     &  1 \rule[-1mm]{0mm}{5mm}\\
fraction &   38 $\pm$ 13 \%& 31 $\pm$ 13 \%&  23 $\pm$ 12 \%&  8 $\pm$ 7 \% 
\rule[-1mm]{0mm}{5mm}\\ \hline
solar type stars & 57 $\pm$ 4 \% & 38 $\pm$ 4 \% & 4 $\pm$ 1 \% & 1 $\pm$ 1 
\%\rule[-1mm]{0mm}{5mm}\\ \hline
\end{tabular} \end{center} \end{table}
\bigskip

\noindent
{\bf \hspace{6mm} $\bullet$\,  Multiplicity and primary spectral type}\medskip

\begin{figure}
\plotfiddle{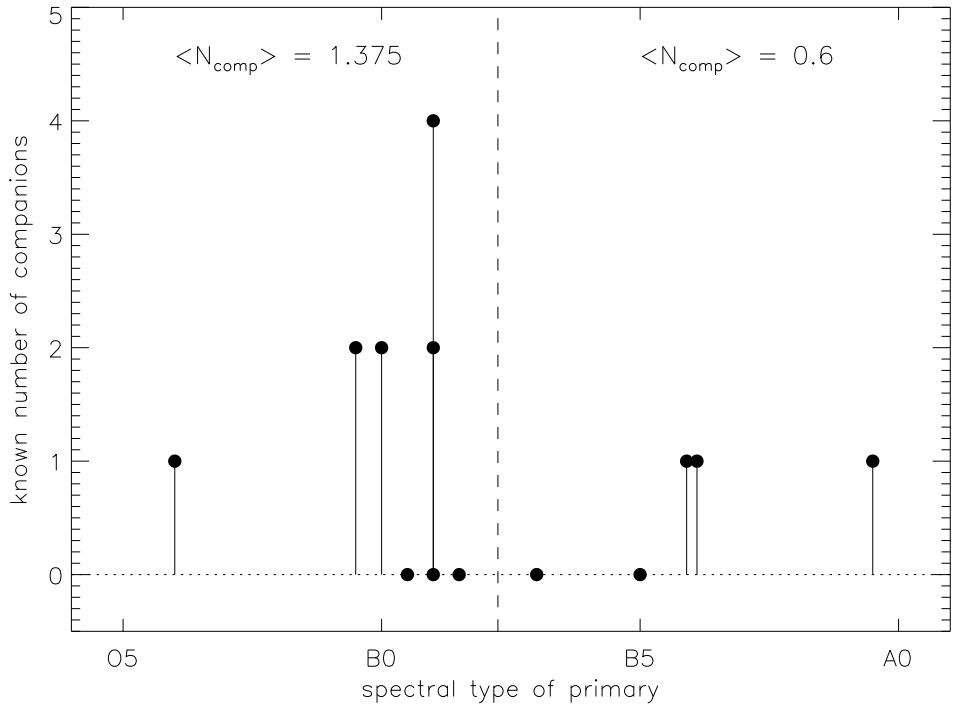}{9.0cm}{0}{80}{80}{-250}{-300}
\caption{The known number of companions is plotted against the
spectral type of the primary.  
}
\end{figure}
\noindent
A remarkable trend for a higher degree of multiplicity among the stars 
of very early spectral type as compared to the later type
stars is apparent in our sample.
The average number of known (visual \& spectroscopic) 
companions per primary
is 2.3 times higher among the primaries with spectral type earlier 
than B3
(11 known companions to 8 primaries) than among the later type  primaries 
(3 known companions to 5 primaries).

\section{Conclusions}

Our results show that
the multiplicity of the massive stars in the Orion Nebula Cluster is
 very high. Even if we only consider the {\bf already known} companions,
there are on average {\bf at least 1.1 companions per high-mass primary star},
about twice as many as found for low-mass primary stars in the general field
as well as in the Orion Nebula Cluster.
If we correct these numbers for observational incompleteness, i.e.~for the
still undetected companions, our data suggest that there should be 
[1.5 \dots 4] \,companions per primary star on average, i.e.~
$\sim$ [3 \dots 8] $\times$ more than  for
low-mass primaries.
Another important difference between the massive stars in the 
Orion Nebula Cluster
and the low-mass stars is the higher
fraction of higher-order multiple systems (triples, quadruplets, \dots) 
among the OB stars.
These findings strongly suggest a different formation mechanism for 
high-mass and low-mass multiple systems.

We also find a trend that 
O- and early B-type stars have more companions than later B-type stars.
The nature of our results seems to support the idea that  high-mass
stars form through collisions of protostars.
However, at least in its present state, this theory provides no 
detailed predictions of the properties of the multiple systems.
If the merging objects would be just spherical stars, one 
would expect most of the multiple systems to have quite small separations, 
probably well
below the resolution limit of the speckle observations ($\sim 20$ AU).
However, the merging protostars are surrounded by extended disks
and/or envelopes, and it is not yet clear how this changes the
outcome of the collision and merging processes.

Furthermore, the dynamical evolution of a cluster will also
affect the properties of the multiple systems.
For example, wide binaries are readily  destroyed by stellar encounters in 
the dense cluster (e.g.~Bonnell 2000). Recent model calculation by
Kroupa et al.~(1999) indicate that, under certain circumstances,
 cluster dynamical evolution can
noticeably change the properties of the multiple systems on time scales 
as short as $\la 1$ Myr. This suggests that even in the very young
Orion Nebula Cluster the observed properties of the multiple systems 
might not necessarily be identical to the primordial multiplicity.

Clearly, further work is necessary for the theoretical as well as the 
observational side of the topic.
As far as observations are concerned, we note that we have already 
performed speckle interferometric observations of the other 14 OB stars
in the Orion nebula cluster. The data analysis will soon be completed
and we should then
be able to draw more firm conclusions on the basis of the
full sample of all 27 OB stars in the Orion Nebula Cluster.

\end{document}